\documentstyle[aps,epsf,amstex]{revtex}

\def\ORDER#1{\hbox{${\cal O}(#1)$}}
\def\NOBAR#1{#1}
\def\BAR#1{\overline{#1}}

\def\KET#1{\vert #1 \rangle}
\def\BRACKET#1#2{\langle #1 \vert #2 \rangle}
\def\hbar{{\mathchar'26\mskip-9muh}}

\begin{document}
\draft
\tighten
\title{Programming Physical Realizations of Quantum Computers}

\author{Hans De Raedt, Kristel Michielsen, and Anthony Hams}

\address{
Institute for Theoretical Physics and Materials Science Centre \\
University of Groningen, Nijenborgh 4 \\
NL-9747 AG Groningen, The Netherlands \\
E-mail: deraedt@@phys.rug.nl, kristel@@phys.rug.nl, A.H.Hams@@phys.rug.nl\\
http://rugth30.phys.rug.nl/compphys
}

\author{Seiji Miyashita and Keiji Saito}

\address{
Department of Applied Physics, School of Engineering \\
University of Tokyo, Bunkyo-ku, Tokyo 113, Japan \\
E-mail: miya@@yuragi.t.u-tokyo.ac.jp, saitoh@@spin.t.u-tokyo.ac.jp
}

\date{\today}
\maketitle
\begin{abstract}
We study effects of the physical realization of quantum computers
on their logical operation.
Through simulation of physical models of quantum computer hardware,
we analyze the difficulties that are encountered in programming
physical realizations of quantum computers.
Examples of logically identical implementations of the controlled-NOT operation
and Grover's database search algorithm are used to demonstrate that
the results of a quantum computation are unstable with respect
to the physical realization of the quantum computer
We discuss the origin of these instabilities and discuss
possibilities to overcome this, for practical purposes,
fundamental limitation of quantum computers.
\end{abstract}

\section{Introduction}\label{sec:1}
Recent theoretical work has shown that a Quantum Computer (QC)
has the potential of solving certain computationally hard problems such as
factoring integers~\cite{SHORone}
and searching databases much faster than a conventional computer~\cite{GROVERone}.
In most theortical work the operation of a QC is described in terms of highly
idealized transformations on the
qubits \cite{DIVINCENZOone,EKERTone,VEDRALone,SHORtwo}.
The impact of the physical implementation of a QC
on its computational efficiency is largely unexplored.

The logical operation of conventional digital circuits
does not depend on their hardware implementation
(e.g. semiconductors, relays, vacuum tubes, etc.).
Dissipative processes suppress the effects of the internal, non-ideal
(chaotic) dynamics and drive the circuits into regions of stable operation.
Conventional digital computers, built from these
digital circuits, are in one particular state at a time and
are able to perform logical operations that do not depend on
their hardware implementation.
From the point of view of programming the computer this is very important.
Implementations of algorithms designed to run on a conventional computer will
give results that do not depend on the hardware used to build the computer.

A QC differs from a conventional digital computer in many respects.
A QC exploits the fact that a quantum system can be in a superposition of states.
Interference of these states allows exponentially
many computations to be done in parallel
\cite{DIVINCENZOone,EKERTone,VEDRALone,SHORtwo,FEYNMAN}.
The presence of the superposition of states is a direct manifestation
of the internal quantum dynamics of the elementary units of the QC, the qubits.
In an ideal QC the qubits are assumed to be ideal two-state quantum systems.
Therefore, the operation of an ideal QC does not depend on the intrinsic
dynamics of its qubits.

A physically realizable QC is a many-body system in which the quantum dynamics
of the qubits is essential to its operation.
Manipulation of one qubit may cause unwanted motion of other qubits.
It is difficult to suppress these effects by dissipation because
in contrast to the case of conventional digital circuits,
dissipation processes have a devastating effect on the coherent quantum
dynamical motion of the qubits.
Therefore a quantum algorithm (QA) 
may yield quantum computation results that depend on the specific physical realization of the QC.
Although QA's can be designed independent of the QC hardware, the implementation of a QA on a
physical realization of a QC (i.e. the programming of the QC) very much depends on the hardware
of which the QC is built from.
We refer to the difficulty of programming QC's as the Quantum Programming Problem (QPP).

Due to the QPP it may be very difficult to develop a non-trivial quantum program for a physical
realization of a QC. Moreover, there is no guarantee that an implementation of a QA,
that works well on one QC will also perform well on other physical realization's of QC's.
As mentioned above, there are several factors that contribute to the QPP:
\begin{itemize}
\item[1)]
Differences between the theoretically perfect
and physically realizable one- and two-qubit operations; qubits
cannot be kept still during the time that other qubits are
being addressed; precision needed to implement operations on the qubits.
\item[2)]
The effect of coupling of the qubits to
other degrees of freedom (dissipation, decoherence).
\end{itemize}

In most theoretical work on QC's and QA's
\cite{SHORone,GROVERone,DIVINCENZOone,EKERTone,VEDRALone,SHORtwo}
one considers theoretically ideal (but physically unrealizable) QC's.
Then the QPP is not an issue.
The QPP is also fundamentally
different from the error propagation previously studied
in QA's implemented on ideal QC's \cite{MIQUEL,PABLO,GUILULONG,BERMANone}
since the QPP is due to the specific realization of the QC
and leads to systematic instead of random errors.

How does a QPP reveal itself?
Consider two logically independent operations ($O_1$ and $O_2$) of the machine.
On a conventional computer or ideal QC, the order
in which we execute these two mutually independent
instructions does not matter: $O_1 O_2 = O_2 O_1$.
However,
on a physically realizable QC {\sl sometimes} the order does matter, even
if there are no logical dependencies in these two program steps.
In some cases, due to practical problems in manipulating
individual qubits $O_1 O_2 \not= O_2 O_1$ and the QC may
give wrong answers.
Note the qualifier {\sl sometimes}. There seems to be
no general rule to decide beforehand which operation
and at what stage of the QA the QPP leads to incorrect results.
At present the only way
to find out seems to be to actually carry out the calculations and
check the results.

In this paper we study the relation
between the physical realization of QC's and their logical operation.
We investigate various aspects of the QPP by simulating QC hardware.
In this work we only consider effects of the intrinsic quantum dynamics of the QC
(item 1, see above).
The study of the effect of the coupling
of the qubits to other degrees of freedom (item 2, see above) is left for future research.
We demonstrate that the programming of a physical, non-ideal implementation of a QC
is difficult, even if the QC consists of only two qubits.
Berman et al. \cite{BERMANtwo} investigated the
influence of the Ising spin interaction on the quantum dynamics of NMR systems.
Although they did not address the QPP, the work is similar in spirit
to \cite{DERAEDT4} and the present paper as
it explores the consequences of the difference
between the ideal QC's and physical realizations of QC's.
For the physical systems and time-scales considered in this paper,
the effects of the interactions between the spins are negligible.
As far as we know no experimental data has been published
that specifically addresses this, for potential applications,
very important and intrinsic problem of programming QC's.
However, with the QC hardware currently available a test of correct
quantum computation on a physical realization of a QC
is definitly within reach.
In this paper we propose two simple QA's that may be used to study the QPP in
physical realizations of QC's.
We also discuss methods to enlarge the region(s) of reliable operation.

The paper is organized as follows:
In Section~\ref{sec:2} we describe a physical model of a QC.
Our choice is largely inspired by NMR-QC experiments
\cite{JONESone,JONEStwo,CHUANGone,CHUANGtwo,MARX,KNILL},
mainly because other candidate technologies
for building QC's
\cite{CIRAC,MONROE,SLEATOR,DOMOKOS,KANE,IMAMO,MAKHLIN,NAKAMURA,NOGUES,MOLMER,SORENSEN,FAZIO,ORLANDO,BLIAS,OLIVEIRA}
are not yet developed to the point that they can execute
computationally non-trivial QA's.
As the basic example of a QA we take the Controlled-NOT (CNOT)
gate \cite{BARENCO}. In Section~\ref{sec:3} we discuss the implementation
of the CNOT gate on an ideal two-qubit QC
and describe simple, non-trivial QA's that consist of repetitions
of CNOT operations.
As an illustration of the general nature of the QPP,
we also consider a more complicated example, namely
Grover's QA to search for an item in a database \cite{GROVERone}.
In Section~\ref{sec:4} we derive the conditions for which the physical two-qubit QC
will exhibit
ideal QC behavior and discuss the generalization of these ideas to $n$-qubit QC's.
Also in Section~\ref{sec:4} we describe the implementation of the QA's discussed in Section~\ref{sec:3}
on a physical realization of a QC.
In Section~\ref{sec:5} we demonstrate and analyze the QPP by simulating the time-evolution of
(= execute QA's on) the physical model of the QC
by solving the time-dependent Schr\"odinger equation.
In Section~\ref{sec:6} we summarize our findings.

\section{Physical Model of a Quantum Computer}\label{sec:2}
Generic QC hardware can be modeled in terms of quantum spins (qubits) that
evolve in time according to the time-dependent Schr\"odinger equation (TDSE)
\begin{equation}
\label{TDSE}
i{\partial \over\partial t} \KET{\Phi(t)}= H(t) \KET{\Phi(t)}
,
\end{equation}
in units such that $\hbar=1$.
For present purposes it is sufficient to consider two-qubit QC's only.
The state
\begin{eqnarray}
\KET{\Phi(t)}&=&
a(\downarrow,\downarrow;t)
\KET{\downarrow,\downarrow}
+a(\uparrow,\downarrow;t)
\KET{\uparrow,\downarrow}
+a(\downarrow,\uparrow;t)
\KET{\downarrow,\uparrow}
+a(\uparrow,\uparrow;t)
\KET{\uparrow,\uparrow}
,
\end{eqnarray}
describes the state of the QC at time $t$.
The complex coefficients
$a(\downarrow,\downarrow;t),\ldots,a(\uparrow,\uparrow;t)$
completely specify the state of the quantum system.
In the absence of interactions with other degrees of freedom
this spin-1/2 system can be modeled by the time-dependent Hamiltonian
\begin{eqnarray}
\label{FULLHAM}
H(t)&=&-J S_1^zS_2^z
-h_{1}^zS_1^{z}-h_{2}^zS_2^{z}
-h_{1}^xS_1^{x}-h_{2}^xS_2^{x}
-h_{1}^yS_1^{y}-h_{2}^yS_2^{y}
\nonumber \\&&
-(\tilde h_{1}^{x}S_1^{x}+ \tilde h_{2}^{x}S_2^{x})\sin (\omega t+\phi_{x})
-(\tilde h_{1}^{y}S_1^{y}+ \tilde h_{2}^{y}S_2^{y})\sin (\omega t+\phi_{y})
,
\end{eqnarray}
where $S_j^{\alpha }$, $\alpha=x,y,z$ denotes the $\alpha$-th component of the spin-1/2
operator representing the $j$-th qubit,
$J$ determines the strength of the interaction between
the two qubits, $h_{j}^\alpha$ and $\tilde h_{j}^\alpha$ represent the strength of
the applied static (magnetic) and applied sinusoidal field (SF) acting on the $j$-th spin respectively.
For a physical system, $h_{2}^\alpha=\gamma h_{1}^\alpha$
and
$\tilde h_{2}^\alpha=\gamma \tilde h_{1}^\alpha$,
for $\alpha=x,y,z$ where $\gamma$ is a constant.
The frequency and the phase of the SF are denoted by $\omega$ and $\phi_{\alpha}$.
As the Ising model, i.e. the first term of (\ref {FULLHAM}), is known to
be a universal QC~\cite{LLOYD1,BERMAN1}, model (\ref {FULLHAM})
is sufficiently general to serve as a physical model for a generic QC at zero temperature.
In terms of spin matrices, the operator $Q_j$ measuring the state of qubit $j$ is given by

\begin{equation}
\label{qubit}
Q_j=\frac{1}{2}-S_j^z.
\end{equation}

For numerical purposes it is necessary to fix as many model parameters as possible.
We have chosen to simulate the two nuclear spins
of the $^1$H and $^{13}$C atoms in a carbon-13 labeled chloroform,
a molecule that has been used in NMR-QC experiments~\cite{CHUANGone,CHUANGtwo}.
In these experiments $h_{1}^z/2\pi \approx 500$MHz, $h_{2}^z/2\pi\approx 125$MHz, and
$J/2\pi\approx -215$Hz~\cite{CHUANGone}.
In the following we will use model parameters rescaled with respect to $h_{1}^z/2\pi$,
i.e we put
\begin{equation}
\label{NMR}
J=-0.43\times10^{-6},\quad h_{1}^z=1, \quad h_{2}^z=0.25.
\end{equation}

Note that there is a difference of many orders of magnitude between the
interaction $J$ and the fields $h_{j}^z$.
If the duration of the SF-pulses is much shorter than $2\pi/|J|$,
the effects of $J$ on the time evolution during these pulses are very small.
Our numerical experiments (see below) are all performed under this condition.
We will only consider QC's at zero temperature without coupling to the environment.
In this sense we simulate highly idealized NMR experiments
on a closed quantum system at zero temperature.
This allows us to study a concrete physical realization
of a QC and at the same time focus on the intrinsic quantum dynamics of the QC.

A QA for QC model (\ref{FULLHAM}) consists of a sequence of
elementary operations (EO) that change the state $\KET{\Psi}$ of the quantum
processor according to the TDSE, i.e. by (a product of) unitary tranformations.
Each EO transforms the input state $\KET{\Psi(t)}$
into the output state $\KET{\Psi(t+\tau)}$ where $\tau$ denotes the
execution time of the EO.
The action of an EO on the state $\KET{\Psi}$
of the quantum processor is defined by specifying
how long it acts (i.e. the time interval $\tau$ during which it is active),
and the values of $J$ and all $h$'s.
During the execution of an EO the values of $J$ and all $h$'s are kept fixed.

The time evolution of quantum model (3) is obtained
by solving TDSE (\ref{TDSE}) for model (\ref{FULLHAM}).
The simulations have been carried out with a
software tool called Quantum Computer Emulator (QCE)\cite{QCEdownload}.
The QCE software simulates physical models of QC hardware
by a Suzuki product-formula~\cite{SUZUKI1,SUZUKI2}, i.e. in terms of elementary unitary
operations~\cite{DERAEDT2,DEVRIES1,DERAEDT3}.
For all practical purposes, the numerical results obtained by this technique are exact.
A detailed description of the QCE software tool can be found elsewhere \cite{DERAEDT1}.

\section{Ideal Quantum Computer}\label{sec:3}
\subsection{Single-qubit operations}

One qubit or one spin-1/2 system is a two-state quantum system. The two basis states
spanning the Hilbert space are denoted by $\KET{\uparrow}\equiv \KET{0}$ and
$\KET{\downarrow}\equiv\KET{1}$.
Rotations of spin $j$ about $\pi/2$ around the $x$ and $y$-axis
are basic QC operations. We will denote them by $X_j$ and $Y_j$ respectively.
In matrix notation, they are given by

\begin{equation}
\label{Xj}
{X_j}\equiv e^{i\pi S^x_j/2}={1\over\sqrt{2}}
\left(
\begin{array}{rr}
1&i \\
i&1
\end{array}
\right)
,
\end{equation}
and
\begin{equation}
\label{Yj}
{Y_j}\equiv e^{i\pi S^y_j/2}={1\over\sqrt{2}}
\left(
\begin{array}{rr}
\phantom{-}1&1 \\ -1&1\\
\end{array}
\right)
.
\end{equation}
Clearly operations such as (\ref{Xj}) and (\ref{Yj}) can be implemented in terms of the time evolution
of model (\ref{FULLHAM}) by a proper choice of the model parameters.
Writing $\KET{a}=a_0\KET{00}+a_1 \KET{10}+a_2\KET{01}+a_3\KET{11}$
with
$\KET{b_1 b_2}\equiv\KET{b_1}\KET{b_2}$ and $b_i=0,1$
we have
\begin{equation}
\label{eq:X1}
X_1\KET{a}=X_1
\left(
\begin{array}{l}
a_0\\ a_1\\ a_2\\ a_3
\end{array}
\right)
={1\over\sqrt{2}}
\left(
\begin{array}{rrrr}
1&i&0&0 \\
i&1&0&0 \\
0&0&1&i \\
0&0&i&1
\end{array}
\right)
\left(
\begin{array}{l}
a_0\\ a_1\\ a_2\\ a_3
\end{array}
\right)
.
\end{equation}
For example, ${\NOBAR X_1}\KET{11}=(\KET{11}+i\KET{01})/\sqrt{2}$.
Using the same labeling of the basis states as in (\ref{eq:X1}) we have

\begin{equation}
\label{eq:Y2}
Y_2
\equiv{1\over\sqrt{2}}
\left(
\begin{array}{rrrr}
1&0&1&0 \\
0&1&0&1 \\
-1&0&\phantom{-}1&\phantom{-}0 \\
0&-1&0&1
\end{array}
\right)
,
\end{equation}
e.g. ${\NOBAR Y_2}\KET{11}=(\KET{10}+\KET{11})/\sqrt{2}$.
The matrix expressions for the inverse of the rotations
$\NOBAR X_1$ and $\NOBAR Y_2$, denoted by $\BAR X_1$
and $\BAR X_2$ respectively, are obtained by taking the hermitian conjugates
of the matrices in (\ref{eq:X1}) and (\ref{eq:Y2}).
For example, ${\BAR Y_2}\KET{11}=(\KET{11}-\KET{10})/\sqrt{2}$.

\subsection{Two-qubit operations: CNOT gate}

Computation requires some form of communication between the qubits.
A basic two-qubit operation is provided by the CNOT gate.
The CNOT gate flips the second spin if the first spin is in the down state, i.e. the first
qubit acts as a control qubit for the second one, see Table \ref{tab:CNOT}.
The procedure that we use to construct the CNOT gate may seem a little ad hoc
and indeed to considerable extent it is.
There is no unique method to construct QC gates.

On an ideal QC the CNOT gate can be implemented by a combination
of single-qubit operations and a two-qubit phase shift operation $P$
defined by the matrix

\begin{equation}
\label{eq:P}
P
\equiv
\left(
\begin{array}{rrrr}
e^{i\phi_{0}}&0&0&0 \\
0&e^{i\phi_{1}}&0&0 \\
0&0&e^{i\phi_{2}}&0 \\
0&0&0&e^{i\phi_{3}}
\end{array}
\right)
.
\end{equation}
Assume that the QC is in a state
\begin{equation}
\KET{\Psi}=a_0\KET{00}+a_1\KET{10}+a_2\KET{01}
+a_3\KET{11}.
\end{equation}
First we apply to $\KET{\Psi}$ the rotation ${\NOBAR Y_2}$, as defined in
(\ref{eq:Y2}). This gives
\begin{eqnarray}
{\NOBAR Y_2}\KET{\Psi}=\frac{1}{\sqrt{2}}[&& (a_0+a_2)\KET{00}+(a_1+a_3)\KET{10}+
(a_2-a_0)\KET{01}
+(a_3-a_1)\KET{11}].
\end{eqnarray}
Next we apply to ${\NOBAR Y_2}\KET{\Psi}$ the phase shift $P$
\begin{eqnarray}
P{\NOBAR Y_2}\KET{\Psi}={1\over\sqrt{2}}
&&\left[ e^{i\phi_{0}}c_0\KET{00}+e^{i\phi_{1}}c_1\KET{10}
+e^{i\phi_{2}}c_2\KET{01}+
e^{i\phi_{3}}c_3\KET{11}\right],
\end{eqnarray}
where $c_0=a_0+a_2$, $c_1=a_1+a_3$, $c_2=a_2-a_0$ and $c_3=a_3-a_1$.
Finally we apply the inverse of the rotation ${\NOBAR Y_2}$
\begin{eqnarray}
\label{eq:full}
{\BAR Y_2}P{\NOBAR Y_2}\KET{\Psi}={1\over 2}
&&\left[
(e^{i\phi_{0}}c_0-e^{i\phi_{2}}c_2)\KET{00}
+(e^{i\phi_{1}}c_1-e^{i\phi_{3}}c_3)\KET{10}
+(e^{i\phi_{0}}c_0+e^{i\phi_{2}}c_2)
\KET{01} 
+(e^{i\phi_{1}}c_1+e^{i\phi_{3}}c_3)
\KET{11}\right]
.
\end{eqnarray}
We now determine the angles $\phi_i$ such that the sequence
(\ref{eq:full}) performs the CNOT operation.
Since the CNOT gate will not change $a_0$ and $a_2$ (see Table \ref{tab:CNOT})
we can choose $\phi_{0}=\phi_{2}$. This gives
\begin{eqnarray}
\label{eq:full1}
{\BAR Y_2}P{\NOBAR Y_2}\KET{\Psi}=e^{i\phi_{0}}
&&\left[ a_0\KET{00} +a_2\KET{01} 
+e^{i\beta } (a_1\cos \alpha
+ia_3\sin \alpha)\KET{10} 
+e^{i\beta}(a_3\cos \alpha
+ia_1\sin \alpha)\KET{11}\right]
,
\end{eqnarray}
where $\beta = \alpha +\phi_{3} -\phi_{0}$ and $\alpha=(\phi_{2}-\phi_{3})/2$.
The global phase factor $e^{i\phi_{0}}$ is physically irrelevant.

The simplest way to implement the phase shift $P$ is to use the
time evolution, i.e. $P=e^{-i\tau H_I}$, of the Ising model
\begin{equation}
\label{Ising}
H_I=-JS_1^zS_2^z -hS_1^z-hS_2^z,
\end{equation}
where the external fields acting on both spins are the same.
From (\ref{Ising}) it follows immediately that
$\phi_{0}=\tau (J/4+h)$, $\phi_{1}=\phi_{2}=-\tau J/4$ and
$\phi_{3}=\tau (J/4 -h)$.
Taking into account our choice $\phi_{0}=\phi_{2}$, (\ref{eq:full1})
becomes
\begin{eqnarray}
{\BAR Y_2}P{\NOBAR Y_2}\KET{\Psi}=e^{i\alpha/2}
&&\left[ a_0\KET{00} +a_2\KET{01} 
+e^{-i\alpha } (a_1\cos \alpha
+ia_3\sin \alpha)\KET{10} 
+e^{-i\alpha }(a_3\cos \alpha
+ia_1\sin \alpha)\KET{11}\right]
.
\end{eqnarray}
Using the same labeling of states as in (\ref{eq:X1}) we have
\begin{equation}
\label{eq:cnotgate}
{\BAR Y_2}P{\NOBAR Y_2}
=
e^{i\alpha /2}
\left(
\begin{array}{cccc}
1&0&0&0 \\
0&e^{-i\alpha}\cos\alpha&0&ie^{-i\alpha}\sin\alpha \\
0&0&1&0 \\
0&ie^{-i\alpha}\sin\alpha&0&e^{i\alpha}\cos\alpha
\end{array}
\right)
.
\end{equation}
Comparing the truth table of the CNOT gate (see Table \ref{tab:CNOT}) with
the matrix in (\ref{eq:cnotgate}), it is clear that putting $\alpha=\pi /2$
will do the job (upto an irrelevant global phase factor).
In terms of Hamiltonian (\ref{Ising}), $-\tau J=\pi$ and $h=-J/2$.
The sequence
\begin{equation}
\label{eq:cnot}
CNOT={\BAR Y_2}I{\NOBAR Y_2}
=e^{i\pi /4}
\left(
\begin{array}{cccc}
1&0&0&0 \\
0&0&0&1 \\
0&0&1&0 \\
0&1&0&0
\end{array}
\right)
,
\end{equation}
performs the CNOT operation on qubit 2 with qubit 1 acting as the control variable.
Here we introduced the symbol $I$ to represent the time evolution $e^{-i\tau H_I}$
with $\tau=-\pi/J$.

\subsection{Quantum Algorithms}

Any QA can be written as a sequence of the one- and two-qubit operations discussed above.
As a simple example of a QA we take $(CNOT)^5$.
On an ideal QC, $CNOT^2$ is the identity operation and hence $(CNOT)^5=CNOT$
but on a physical QC this is not always the case, see below.
To illustrate the dependence of the quantum computation on
the physical implementation and on the choice of the input state
we consider two QA's, $QA_1$ and $QA_2$, defined by

\begin{eqnarray}
\label{eq:qa1}
QA_1\KET{b_1b_2}&\equiv&
(CNOT)^5\KET{b_1b_2},
\\
\label{eq:qa2}
QA_2\KET{singlet}
&\equiv&
Y_1 (CNOT)^5\KET{singlet},
\end{eqnarray}%
where $\KET{singlet}=(\KET{01}-\KET{10})/\sqrt{2}$.
We have

\begin{eqnarray}
\label{eq:qa3}
(\KET{01}-\KET{11})/\sqrt{2}&=&(CNOT)^5\KET{singlet},
\end{eqnarray}%
and hence $\BRACKET{singlet}{(CNOT)^5Q_1(CNOT)^5|singlet}=1/2$.
We can obtain a clear-cut answer in terms of expectation values of the qubits
by applying a $\pi/2$ rotation of spin 1

\begin{eqnarray}
\label{eq:qa4}
\KET{11}&=&Y_1(CNOT)^5\KET{singlet}.
\end{eqnarray}%
Therefore in (\ref{eq:qa2}), the CNOT operations are followed
by a $\pi/2$ rotation of spin 1.

As a more complicated example of a QA, we consider Grover's database search algorithm
to find the needle in a haystack.
On a conventional computer, finding an item out of
of $N$ elements requires \ORDER{N} queries\cite{CORMEN}.
Grover has shown that a QC can find the item using
only \ORDER{\sqrt{N}} attempts\cite{GROVERone}.
The reduction from \ORDER{N} to \ORDER{\sqrt{N}} is due to
the intrinsic massive parallel operation of the QC.
Assuming a uniform probability distribution for the needle,
for $N=4$ the average number of queries required by a conventional
algorithm is 9/4\cite{CHUANGone,CORMEN}.
With Grover's QA the correct answer can be found in
a single query\cite{JONESone,CHUANGone}.

Experimentally Grover's QA has been implemented on a 2-qubit NMR-QC for
the case of a database containing four items\cite{JONESone,CHUANGone}.
In experiments\cite{JONESone,CHUANGone} the sequences

\begin{eqnarray}
\label{eq:Grover0}
U_0 &=&{     X}_1{\BAR Y}_1 {     X}_2{\BAR Y}_2 G
     {     X}_1 {\BAR Y}_1 {     X}_2 {\BAR Y}_2 G
{\BAR X}_1 {\BAR X}_1 {\BAR Y}_1
{\BAR X}_2 {\BAR X}_2 {\BAR Y}_2
     , \\
\label{eq:Grover1}
U_1 &=&{     X}_1{\BAR Y}_1 {     X}_2{\BAR Y}_2 G
{     X}_1 {\BAR Y}_1 {\BAR X}_2 {\BAR Y}_2 G
{\BAR X}_1 {\BAR X}_1 {\BAR Y}_1
{\BAR X}_2 {\BAR X}_2 {\BAR Y}_2
,  \\
\label{eq:Grover2}
U_2 &=&{     X}_1{\BAR Y}_1 {     X}_2{\BAR Y}_2 G
{\BAR X}_1 {\BAR Y}_1 {     X}_2 {\BAR Y}_2 G
{\BAR X}_1 {\BAR X}_1 {\BAR Y}_1
{\BAR X}_2 {\BAR X}_2 {\BAR Y}_2
, \\
\label{eq:Grover3}
U_3 &=&{     X}_1{\BAR Y}_1 {     X}_2{\BAR Y}_2 G
{\BAR X}_1 {\BAR Y}_1 {\BAR X}_2 {\BAR Y}_2 G
{\BAR X}_1 {\BAR X}_1 {\BAR Y}_1
{\BAR X}_2 {\BAR X}_2 {\BAR Y}_2
,
\end{eqnarray}%
have been chosen to implement Grover's search algorithm.
The subscript $j$ of $U_j$ corresponds to the position
of the searched-for item in the database.
In all four cases the input state is $\KET{00}$.
The two-qubit operation $G$ is defined by

\begin{equation}
\label{eq:G}
G=
\left(
\begin{array}{cccc}
e^{-i\pi/4}&0&0&0 \\
0&e^{+i\pi/4}&0&0 \\
0&0&e^{+i\pi/4}&0 \\
0&0&0&e^{-i\pi/4}
\end{array}
\right)
,
\end{equation}%
and performs a conditional phase shift.

On an ideal QC the QA's (\ref{eq:Grover0}) -- (\ref{eq:Grover3}) are by no means unique:
Various alternative expressions can be written down by using the algebraic
properties of the $X$'s and $Y$'s. This feature has been exploited to eliminate redundant
elementary operations\cite{CHUANGone}.
On an ideal QC sequences
(\ref{eq:Grover0}) -- (\ref{eq:Grover3})
return the correct answer, i.e. the position of the searched-for item.
This is easily verified on the QCE by selecting the elementary operations
that implement an ideal QC.

\section{Physical Quantum Computer}\label{sec:4}
\subsection{Single-qubit operations}

NMR uses SF pulses to rotate the spins. By tuning the
frequency of the SF to the precession frequency of a particular spin ($h_{j}^z$
in our case),
the power of the applied pulse (= intensity times duration) controls how much the spin will rotate.
The axis of the rotation is determined by the direction of the applied SF.
The elementary model of an NMR experiment on a single spin (qubit 1 for example) subject to
a constant magnetic field along the $z$-axis and a SF along the $x$-axis reads~\cite{BAYM}
\begin{equation}
\label{NMRTDSE}
i{\partial\over\partial t}\KET{\Phi(t)}=
-\left( h_{1}^z S_1^z +\tilde h_{1}^x S_1^x \sin \omega t \right) \KET{\Phi(t)},
\end{equation}
where
$\KET{\Phi} =\KET{\uparrow}\BRACKET{\Phi}{\uparrow} +\KET{\downarrow}\BRACKET{\Phi}{\downarrow}$,
$\KET{\Phi(t=0)}$ is the initial state of the two-state system
and we have set the phase $\phi_{x}=0$.
Substituting $\KET{\Phi(t)}=e^{ith_{1}^z S_1^z}\KET{\Psi(t)}$ yields
\begin{eqnarray}
i{\partial\over\partial t}\KET{\Psi(t)}=
-\tilde h_{1}^x&& \left(
S_1^x \sin \omega t\cos h_{1}^z t 
+ S_1^y \sin \omega t\sin h_{1}^z t
\right) \KET{\Psi(t)}.
\end{eqnarray}
At resonance, i.e. $\omega=h_{1}^z$, we find
\begin{eqnarray}
\label{TDSENMR1}
i{\partial\over\partial t}\KET{\Psi(t)}=
-{\tilde h_{1}^x\over2}&& \left(  S_1^y  + S_1^x \sin 2\omega t 
- S_1^y \cos 2\omega t
\right) \KET{\Psi(t)}.
\end{eqnarray}
Assuming that the effects of the higher harmonic terms
(i.e. the terms in $\sin 2\omega t$ and $\cos 2\omega t$)
are small~\cite{BAYM}, (\ref{TDSENMR1}) is easily solved to give
\begin{equation}
\KET{\Psi(t)}\approx e^{i t \tilde h_{1}^x S_1^y/2}
\KET{\Psi(t=0)},
\end{equation}
so that the overall action of a SF-pulse of duration $\tau$
can be written as
\begin{equation}
\label{ROTNMR}
\KET{\Phi(t+\tau)}\approx e^{i\tau h_{1}^z S_1^z} e^{i \tau \tilde h_{1}^x S_1^y/2}
\KET{\Phi(t)}.
\end{equation}
Hence it follows that application of an SF-pulse of power
$\tau \tilde h_{1}^x=\pi$ will have the effect of rotating spin 1 by
an angle of $\pi/2$ about the $y$-axis, as is clear by comparing (\ref{Yj}) with (\ref{ROTNMR}).

In deriving (\ref{ROTNMR}), higher harmonics have been neglected, as indicated
by the ``$\approx$'' sign.
Instead of applying SF's along the $x$ or $y$ direction,
one may also consider using SF's that rotate in the $x$-$y$ plane.
This leads to the TDSE \cite{SLICHTER}

\begin{equation}
\label{NMRRFTDSE}
i{\partial\over\partial t}\KET{\Phi(t)}=
-\left[ h_{1}^z S_1^z +
\tilde h_{1}^x (S_1^x \sin \omega t + S_1^y \cos \omega t) \right] \KET{\Phi(t)},
\end{equation}
and instead of (\ref{ROTNMR}) we obtain
\begin{equation}
\label{ROTRF}
\KET{\Phi(t+\tau)}= e^{i\tau h_{1}^z S_1^z} e^{i \tau \tilde h_{1}^x S_1^y}
\KET{\Phi(t)}.
\end{equation}

A QC contains at least two spins.
If in experiments it is difficult to shield a particular spin from the SF,
an application of an SF pulse affects not only the state of the resonant spin
but changes the state of the other spins too
(unless they are perfectly aligned along the $z$-axis).
A general analytical, quantitative analysis of this many-body problem is rather difficult.
We will study the limiting case in which
the interaction between the spins has neglible impact on the time evolution of the spins
during application of the SF pulse.
As our numerical results (see below) demonstrate, this is the case that is relevant to
the model system considered in the present paper
and also to experiments \cite{JONESone,JONEStwo,CHUANGone,CHUANGtwo}.

We consider the two-spin system described by the TDSE

\begin{equation}
\label{RFTDSE1}
i{\partial\over\partial t}\KET{\Phi(t)}=
-\left[
h_{1}^z S_1^z +h_{2}^z S_2^z
+\tilde h_{1}^x( S_1^x \sin \omega t +S_1^y \cos \omega t)
+\tilde h_{2}^x( S_2^x \sin \omega t +S_2^y \cos \omega t)
\right]
 \KET{\Phi(t)}.
\end{equation}
Substituting $\KET{\Phi(t)}=e^{it\omega( S_1^z+S_2^z)}\KET{\Psi(t)}$ we obtain

\begin{equation}
\label{RFTDSE2}
i{\partial\over\partial t}\KET{\Psi(t)}=
-\left[
(h_{1}^z -\omega)S_1^z +(h_{2}^z-\omega) S_2^z
+\tilde h_{1}^xS_1^y +\tilde h_{2}^x S_2^y
\right]
 \KET{\Psi(t)}.
\end{equation}
Our aim is to rotate spin 1 about an angle $\varphi_1$ without affecting
the state of spin 2. This can be accomplished as follows.
First we choose

\begin{equation}
\label{CONDITION0}
\omega=h_{1}^z,
\end{equation}%
i.e. the frequency of the SF pulse is tuned to the resonance frequency of spin 1.
Then (\ref{RFTDSE2}) can easily be integrated. The result is

\begin{equation}
\label{RFTDSE3}
\KET{\Phi(t)}=
e^{ith_{1}^z( S_1^z+S_2^z)}
e^{it\tilde h_{1}^x S_1^y}
e^{it{\bf S}_2\cdot{\bf v}_{1,2}}
 \KET{\Phi(0)},
\end{equation}%
where ${\bf v}_{n,m}\equiv(0,\tilde h_{m}^x,h_{m}^z - h_{n}^z )$.

The third factor in (\ref{RFTDSE3}) rotates spin 2 around the vector ${\bf v}_{1,2}$.
This factor can be expressed as

\begin{equation}
\label{RFTDSE4}
e^{it{\bf S}_m\cdot{\bf v}_{n,m}}
=
\left(
\begin{array}{cc}
1&0 \\
0&1 \\
\end{array}
\right)
\cos \frac{t|{\bf v}_{n,m}|}{2}
+
i|{\bf v}_{n,m}|^{-1}\left(
\begin{array}{cc}
h_{m}^z - h_{n}^z&-ih_{m}^x \\
ih_{m}^x&h_{n}^z - h_{m}^z \\
\end{array}
\right)
\sin \frac{t|{\bf v}_{n,m}|}{2},
\end{equation}%
and we see that the SF pulse will not change the state of spin 2
if and only if the duration $t_1$ of the pulse satisfies

\begin{equation}
\label{CONDITION3}
t_1|{\bf v}_{1,2}|=t_1\sqrt{(h_{1}^z - h_{2}^z)^2+(\tilde h_{2}^x)^2}=4\pi n_1,
\end{equation}%
where $n_1$ is a positive integer.

The second factor in (\ref{RFTDSE3}) is a special case of (\ref{RFTDSE4}).
It is easy to see that if

\begin{equation}
\label{CONDITION2}
t_1\tilde h_{1}^x=\varphi_1,
\end{equation}%
the second factor in (\ref{RFTDSE3})
will rotate spin 1 about $\varphi_1$ around the $y$-axis.
Therefore, if conditions
(\ref{CONDITION0}),
(\ref{CONDITION3}), and
(\ref{CONDITION2})
are satisfied we can rotate spin 1 about $\varphi_1$
without affecting the state of spin 2,
independent of the physical realization of the QC.
However, the first factor in (\ref{RFTDSE3}) can still
generate a global phase shift.
Although it drops out of the expression
of the expectation value of the qubits, in general
it has to be taken into account in a QC calculation
because this phase shift depends on the state of the spins.
Adding the condition

\begin{equation}
\label{CONDITION1}
t_1h_{1}^z=4\pi k_1,
\end{equation}%
where $k_1$ is a positive integer ($h_{i}^z>0$ by definition),
the first factor in (\ref{RFTDSE3}) is always equal to one.
Summarizing: If conditions
(\ref{CONDITION0}),
(\ref{CONDITION3}),
(\ref{CONDITION2}), and
(\ref{CONDITION1})
are satisfied we can rotate spin 1 about $\varphi_1$
without affecting the state of spin 2 and
without introducing unwanted phase shifts.

A last constraint on the choice of the pulse parameters comes
from the fact that

\begin{equation}
\label{CONDTION4}
h_{2}^\alpha=\gamma h_{1}^\alpha\quad,\quad
\tilde h_{2}^\alpha=\gamma\tilde h_{1}^\alpha\quad;\quad  \alpha=x,y,z.
\end{equation}%
Without loss of generality we will assume that $0<\gamma<1$, in concert
with the choice of parameters (\ref{NMR}).

Using constraint (\ref{CONDTION4}) and conditions
(\ref{CONDITION0}),
(\ref{CONDITION3}),
(\ref{CONDITION2}), and
(\ref{CONDITION1})
we have

\begin{eqnarray}
\label{CONDITIONa}
(1-\gamma)^2 k_1^2+\frac{\gamma^2}{4}\left(\frac{\varphi_1}{2\pi}\right)^2=n_1^2,
\end{eqnarray}%
and reversing the role of spin 1 and spin 2 we obtain

\begin{eqnarray}
\label{CONDITIONb}
(1-\frac{1}{\gamma})^2 k_2^2+\frac{1}{4\gamma^2}\left(\frac{\varphi_2}{2\pi}\right)^2=n_2^2,
\end{eqnarray}%
where $k_1$, $k_2$, $n_1$, and $n_2$ are positive integers.
The angles of rotation about the $y$-axis can be chosen such that
$0\le \varphi_1\le 2\pi$ and $0\le \varphi_2\le 2\pi$.

In general (\ref{CONDITIONa}) or (\ref{CONDITIONb}) have no solution but a good approximate
solution may be obtained if
$\gamma$ is a rational number and $k_1$ and $k_2$ are large.
Let  $\gamma=N/M$ where $N$ and $M$ are integers satisfying $0<N<M$.
It follows that
the representation $k_1=kMN^2$ and $k_2=kNM^2$ will generate
sufficiently accurate solutions of (\ref{CONDITIONa}) and (\ref{CONDITIONb})
if the integer $k$ is chosen such that

\begin{equation}
\label{CONDITIONk}
2kNM(M-N)\gg1.
\end{equation}%
In terms of $k$, $N$, and $M$, the relevant physical quantities are then given by

\begin{equation}
\label{PARAMETERS1}
\frac{t_1h_{1}^z}{2\pi}=2kMN^2
\quad,\quad
\frac{\tilde h_{1}^x}{h_{1}^z}=\frac{1}{2kMN^2}\frac{\varphi_1}{2\pi},
\end{equation}%
and

\begin{equation}
\label{PARAMETERS2}
\frac{t_2h_{1}^z}{2\pi}=2kM^3\quad,\quad
\frac{\tilde h_{2}^x}{h_{1}^z}=\frac{1}{2kM^3}\frac{\varphi_2}{2\pi}.
\end{equation}%
In our numerical experiments we use
(\ref{PARAMETERS1}) and (\ref{PARAMETERS2})
to determine
the duration of the SF pulses for both the static and rotating SF's.
In the latter case the SF pulses will be optimized in the sense
that a pulse that rotates spin 1 (2) will hardly affect spin 2 (1)
if $k$ satisfies condition (\ref{CONDITIONk}).

The assumption of a pure sinusoidal time dependence of the applied fields
serves to simplify the analytical analysis given above.
In experiment there is no good reason to stick to a simple time dependence of the pulses\cite{FREEMAN}.
In general

\begin{equation}
\label{WTDSE1}
i{\partial\over\partial t}\KET{\Phi(t)}=
-\left[
h_{1}^z S_1^z +h_{2}^z S_2^z
+w(t)\tilde h_{1}^x( S_1^x \sin \omega t +S_1^y \cos \omega t)
+w(t)\tilde h_{2}^x( S_2^x \sin \omega t +S_2^y \cos \omega t)
\right]
 \KET{\Phi(t)},
\end{equation}
where $w(t)$ can be almost any waveform.
For $\omega=h_1^z$, the formal solution of (\ref{WTDSE1}) reads

\begin{equation}
\label{WTDSE2}
\KET{\Phi(t)}=
e^{ith_{1}^z( S_1^z+S_2^z)}
\exp\left(i\int_0^t du\,w(u)\tilde h_{1}^xS_1^y\right)
\exp_+\left\{i\int_0^t du\,\left[ (h_2^z-h_1^z)S^z_2+w(u)\tilde h_{2}^xS_2^y\right]\right\}
 \KET{\Phi(0)},
\end{equation}%
where $\exp_+\{\ldots\}$ denotes the time-ordered exponential.
The introduction of a general form of $w(t)$ replaces condition
(\ref{CONDITION3})
by

\begin{equation}
\label{WCONDITION3}
\exp_+\left\{i\int_0^{t_1} du\,\left[ (h_2^z-h_1^z)S^z_2+w(u)\tilde h_{2}^xS_2^y\right]\right\}=1,
\end{equation}%
and condition (\ref{CONDITION2}) becomes

\begin{equation}
\label{WCONDITION2}
t_1\tilde h_{1}^x\int_0^{t_1} du\,w(u)=\varphi_1,
\end{equation}%
expressing the fact that the rotation angle $\varphi_1$ is determined
by the power of the pulse only.
Conditions (\ref{CONDITION0}) and (\ref{CONDITION1}) remain the same.
There are many forms of $w(u)$ that will satisfy (\ref{WCONDITION2}),
so in this respect there is a lot of freedom in the choice of $w(t)$.
Finding the form of $w(u)$ such that also (\ref{WCONDITION3}) holds
is a (complicated) optimization problem,
in particular when the QC contains several qubits.

To summarize: If conditions
(\ref{CONDITION0}),
(\ref{CONDITION3}),
(\ref{CONDITION2}), and
(\ref{CONDITION1})
are satisfied we can rotate spin 1 about $\varphi_1$
without affecting the state of spin 2
and
without introducing unwanted phase shifts.
In practice we may replace
(\ref{CONDITION3}) and
(\ref{CONDITION2}), by
(\ref{WCONDITION3}) and
(\ref{WCONDITION2}) respectively.

For our choice (\ref{NMR}) of the model parameters, $\gamma=1/4$
such that $N=1$ and $M=4$. In general $\gamma$ will
not be a ratio of two small integers but it may be approximated
to any desired precision by a rational number.
Let us consider the hypothetical case ($N=11$, $M=40$) such
that $\gamma=11/40=0.275$.
Then (\ref{CONDITIONk}) reads $25520k\gg1$ so that the choice
$k=1$ already yields an accurate solution to
(\ref{CONDITIONa}) and
(\ref{CONDITIONb}).
However as $t_1h_{1}^z/2\pi=9680$ and $t_2h_{1}^z/2\pi=128000$,
rather long SF pulses are required to perform these
nearly ideal, single-qubit operations.
As this example shows, the duration of the pulses
that implement accurate single-qubit operations will be determined by
the representation of $\gamma$ as a ratio of two (small) integers.

From (\ref{PARAMETERS1}) and (\ref{PARAMETERS2}) it follows that
$|\tilde h^x_j|\ll|h_1^z-h_2^z|$ for accurate single-qubit operations.
This implies $t_1h^z_1(1-\gamma)=4\pi n_1$ or $k_1(M-N)=Mn_1$.
For $N=1$ and $M=4$ we see that $k_1$ has to be a multiple of $4$.
This reasoning readily generalizes: For $n$ spins
$k_1(M_j-N_j)=M_jn_j$ for integer $n_j$ and $j=2,\ldots,n$.
In other words, the frequencies of precession of each of the qubits
have to be commensurate with each other. Otherwise systematic phase errors
will be generated in the course of the computation.
This conclusion does not depend on the peculiarities of
the NMR technique: It holds in general.

\subsection{Two-qubit operations: CNOT gate}

As the CNOT sequence (\ref{eq:cnot}) has been constructed on the basis
of model (\ref{Ising}), some modification is necessary to account
for the fact that the two nuclear spins feel different static
fields (see (\ref{FULLHAM})).
In general the Hamiltonian reads
\begin{equation}
\label{eq:HNMR}
H_{NMR}=-J S_1^zS_2^z-h_{1}^zS_1^{z}-h_{2}^zS_2^{z}
.
\end{equation}
Comparison of (\ref{Ising}) with (\ref{eq:HNMR})
shows that the implemention of the CNOT operation
requires additional rotations:

\begin{eqnarray}
\label{eq:cnotnmr}
CNOT&=&
{\BAR Y_2}
e^{-i\tau(h_{1}^z-h)S_1^{z}}
e^{-i\tau(h_{2}^z-h)S_2^{z}}
e^{-i\tau H_{NMR}}{\NOBAR Y_2},
\nonumber \\
&=&
{\BAR Y_2}
e^{-i\tau(h_{1}^z-h)S_1^{z}}
e^{-i\tau(h_{2}^z-h)S_2^{z}}
{\NOBAR Y_2}
{\BAR Y_2}
e^{-i\tau H_{NMR}}{\NOBAR Y_2}
,
\end{eqnarray}
where we used the fact that ${\NOBAR Y_2}{\BAR Y_2}=1$.
The extra phase shifts in (\ref{eq:cnotnmr}) can be expressed
in terms of single-qubit operations. The idendities
\begin{eqnarray}
\label{eq:cnotnmr1}
e^{-i\tau(h_{1}^z-h)S_1^{z}}
&=&
{\NOBAR Y_1} {\NOBAR X_1^\prime} {\BAR Y_1}
=
{\BAR X_1} {\NOBAR Y_1^\prime} {\NOBAR X_1},
\\
\label{eq:cnotnmr2}
e^{-i\tau(h_{2}^z-h)S_2^{z}}
&=&
{\NOBAR Y_2} {\NOBAR X_2^\prime} {\BAR Y_2},
\end{eqnarray}
define the single-spin rotations
${\NOBAR X_1^\prime}$,
${\NOBAR Y_1^\prime}$,
and
${\NOBAR X_2^\prime}$.

As (\ref{eq:cnotnmr1}) and (\ref{eq:cnotnmr2}) suggest,
there are many different, logically equivalent
sequences that implement the CNOT gate on an NMR QC. We have chosen to limit ourselves
to the respresentations

\begin{eqnarray}
\label{eq:cnot1}
CNOT_{1}
&=&
{\NOBAR Y_1}{\NOBAR X_1^\prime}{\BAR Y_1}{\NOBAR X_2^\prime}{\BAR Y_2}{I^\prime}{\NOBAR Y_2},
\\
\label{eq:cnot2}
CNOT_{2}
&=&
{\NOBAR Y_1}{\NOBAR X_1^\prime}{\NOBAR X_2^\prime}{\BAR Y_1}{\BAR Y_2}{I^\prime}{\NOBAR Y_2},
\\
\label{eq:cnot3}
CNOT_{3}
&=&
{\BAR X_1}{\NOBAR Y_1^\prime}{\NOBAR X_2^\prime}{\BAR Y_2}{\NOBAR X_1}{I^\prime}{\NOBAR Y_2},
\end{eqnarray}
where we introduced the symbol $I^\prime$ to represent the time evolution $e^{-i\tau H_{NMR}}$
with $\tau=-\pi/J$.
On an ideal QC the sequences (\ref{eq:cnot1}) -- (\ref{eq:cnot3}) give identical results.
On an NMR-like QC they do not because operations such as $X_1$ and $X_2$ no longer commute.
We will use sequences (\ref{eq:cnot1}) -- (\ref{eq:cnot3}) to demonstrate the QPP.

\subsection{Quantum Algorithms}

On a conventional computer an algorithm is a sequence of logical operations that
defines a one-to-one relation between the input and output data.
We expect that a conventional computer always returns
the correct result, irrespective of the input.
Also a QC should have correct (input, output) relationships.
In contrast to a conventional computer, a QC accepts
as input linear superpositions of basis states and
can return superpositions as well.
If a quantum gate correctly operates on each of the basis states,
it will also do so on any general linear superposition
unless the operation generates additional phase factors that
depend on the input state.
Of course this does not happen on an ideal QC but on
a realistic one it may.
Above we have shown how to reduce unwanted phase errors that result from
imperfections of the one- and two-qubit operations.

For each realization of QC hardware, there is a one-to-one correspondence between
the QA and the unitary matrix that transforms the superposition accordingly.
A QA will operate correctly under {\sl all} circumstances if
the {\sl whole} unitary matrix representing the QA is
a good approximation to the ideal one.
In other words, the magnitude and the phase of {\sl all} matrix elements
should be close to their ideal values.
It is not sufficient to have for example two different CNOT gates that operate
correctly by themselves: Also the relative phases that they produce should match.
For $n$ qubits there are $2^n(2^n-1)$ real numbers that specify the unitary matrix
corresponding to a QA. All these numbers should be close to their ideal values,
otherwise the QA is bound to produce wrong answers.

Experimental realizations of QC's have not yet demonstrated that
a QC can correctly compute the answer for inputs other than simple basis states.
However, with the QC hardware currently available such a test
is definitly within reach.
The two simple QA's, (\ref{eq:qa1}) and (\ref{eq:qa2}) may be used for this purpose.

In general on a physical QC, $CNOT^2\not=1$ and hence $(CNOT)^5$ in
(\ref{eq:qa1}) and (\ref{eq:qa2}) does not reduce to one
CNOT operation.
The effect of the physical implemention of a QC
on the logical operation of a QA
will be most clear if we can distinguish errors
due to faulty input data from those that are intrinsic to the
physics of the qubits.
Therefore we will provide the exact input state to the QA and compare
the result returned by the QA with the exact answer.
This procedure simplifies the analysis but does not touch the essence of the matter.
We prepare $\KET{b_1b_2}$ and $\KET{singlet}=(\KET{01}-\KET{10})/\sqrt{2}$
by starting from the state $\KET{00}$ and by performing exact rotations of the spins.

In the case of Grover's database search algorithm,
the representation of $G$ in terms of the time evolution of (\ref{eq:HNMR}) reads

\begin{eqnarray}
\label{eq:Gnmr}
G&=&e^{-i\pi S_1S_2}=e^{-i\tau h_{1}^z S_1^{z}}
e^{-i\tau h_{2}^zS_2^{z}} e^{-i\tau H_{NMR}}
=
{\NOBAR Y_2} {\NOBAR X_2^{\prime\prime}} {\BAR Y_2}
{\NOBAR Y_1} {\NOBAR X_1^{\prime\prime}} {\BAR Y_1}
e^{-i\tau H_{NMR}}
,
\end{eqnarray}
where $\tau =-\pi/J$.
This choice of $\tau$ also fixes the angles the rotations,
and
through relations (\ref{PARAMETERS1}) and (\ref{PARAMETERS2})
also all parameters of the operations
${\NOBAR X_1^{\prime\prime}}$ and ${\NOBAR X_2^{\prime\prime}}$.

\section{Simulation}\label{sec:5}
\subsection{Model parameters}

The parameters of model (\ref{FULLHAM}) for which
$e^{-i\tau H}$ implements the EO's of the ideal QC are listed
in Table \ref{tab:IDEALQC}.
On an NMR-like QC, the one-qubit operations can be implemented
by applying SF pulses, as explained above. The two-qubit
operation $I^\prime$ can be implemented by letting the system
evolve in time according to Hamiltonian $H_{NMR}$, given by (\ref{eq:HNMR}).
$I^\prime$ is the same for both an ideal or NMR-like QC.
Note that the condition $\tau J=-\pi$ yields
$\tau/2\pi=1162790.6977$, a fairly large number
(compared to our reference $h_{1}^z=1$, see (\ref{NMR})).

The model parameters for the fixed and rotating SF's are determined according
to the theory outlined above. We use the integer $k$ to compute all free
parameters and the subscript $s=2kMN^2$ to label the results of the QC calculation.
For reference we present the set of parameters corresponding to $k=1$
for QC's using fixed and rotating SF in Tables \ref{tab:NMRQC} and \ref{tab:NMRRFQC} respectively.

\subsection{Results}

As a standard test we execute all sequences on an implementation of the ideal QC
(see Table \ref{tab:IDEALQC}). They all give the exact answers (results not shown).
Furthermore the results (not shown) do not change if we put $J=0$ in all single-qubit
operations, which is not a surprise in view of the fact that typical pulse durations are
much smaller than $1/2|J|$.
It is also necessary to rule out that the numerical results depend on the
time step $\delta$ used to solve the TDSE.
The numerical error of the product formula used by QCE is proportional to $\delta^2$
\cite{DERAEDT2,DEVRIES1,DERAEDT3}.
It goes down by a factor of about one hundred if we reduce the time step by a factor of 10.
Within the two-digit accuracy used to present our data,
there is no difference between the results for $\delta=0.01$ and $\delta=0.001$.
Hence we can be confident that we are solving the TDSE with a sufficiently high accuracy.

In Table \ref{tab:NMRRFRESULT1} we present simulation results for QA's, $QA_1$ and
$QA_2$ defined by (\ref{eq:qa1}) and (\ref{eq:qa2}) respectively.
It is clear that even the least accurate implementation ($s=8$, $k=1$)
nicely reproduces the correct answers if the input corresponds to
one of the four basis states.
The corresponding entries for $QA_2$ seem to suggest that $CNOT_1$ is working well for $s=8$.
However the result for $s=16$ ($k=2$) shows that the apparently
good result for $s=8$ is accidental, as we might have expected on the basis
of criterion (\ref{CONDITIONk}) (which in this case reads $24\gg1$).
In agreement with the theoretical analysis of Section~\ref{sec:4}.A
the results converge to the exact ones for sufficiently large $k$, as
shown in Table \ref{tab:NMRRFRESULT1}.
For small $s$, the difference in the accuracy with which $QA_1$ and $QA_2$
give the correct answer clearly shows that in order for a QA to work
properly, it is not sufficient to show that it correctly operates
only when the input corresponds to one of the basis states.

In the regime where phase errors are significant the QA's exhibit
the QPP. This is exemplified in Tables
\ref{tab:NMRRFRESULT2}
and
\ref{tab:NMRRFRESULT3}
where we show the results of using $CNOT_2$ and $CNOT_3$ instead of $CNOT_1$.
For $k<32$ there is a clear signature of the QPP:
These logically identical QA's are sensitive
to the order in which the single-qubit operations are carried out.

The results presented in Tables \ref{tab:NMRRFRESULT1}--\ref{tab:NMRRFRESULT3}
have been obtained using rotating SF's. As explained above, in this case
a single-qubit operation on qubit $j$ {\sl exactly} rotates qubit $j$
about the specified angle (but perturbs the state of the other spin).
In Table \ref{tab:NMRRESULT1} we present simulation results
obtained by using SF in the $x$ or $y$ direction only. Then
the single-spin rotation on spin $j$ no longer corresponds
to the exact one. Nevertheless, as Table \ref{tab:NMRRESULT1} shows,
for sufficiently large $s$ the results nicely converge
to the correct anwers. Apparently, for a QA to compute correctly,
it is more important to have the phase errors under control than
to perform very accurate single-spin rotations.

The very essence of QA's is the use of entangled states at some
stage of the calculations. It is at this point that the QA
is most sensitive to (accumulated) phase errors.
As another illustration of this phenomenon, we present
in Tables
\ref{tab:GROVRESULT1}
and
\ref{tab:GROVRESULT2}
some typical results obtained by excuting Grover's database search
algorithm on the same model QC's as those used in the examples
discussed above. We find that reasonably good answers are
obtained if $s\ge32$, in concert with the observations based on
QA's $QA_1$ and $QA_2$.

The results discussed above show effects of imperfections in the physical
implementation of single-qubit operations.
Thereby we assumed that $J$, and the static applied fields
$h^z_1$ and $h^z_2$ are fixed in time and known to very high precision.
The Ising-model time evolution was used to perform two-qubit operations,
leaving only the duration of this operation as a possible source for
causing errors.
In Table \ref{tab:NMRRFRESULT4} we give examples
of the extreme sensitivity of a QA to the precision with
which the parameters have to be specified.
Essentially we repeated the calculation of Table \ref{tab:NMRRFRESULT1} for $s=256$
but on purpose we made an error in the specification of the duration of
$I^{\prime}$. As Table \ref{tab:NMRRFRESULT4} shows, an error
in the 8-th digit can have a devastating effect on the outcome
of the QC calculation.
This again is just another manifestation of the QPP but not really a surprise:
During the application of $I^{\prime}$ the spins
rotate around the $z$-axis with their resonance frequencies $h_1^z$
and $h_2^z$. A small deviation in $\tau/2\pi$ from its ideal value
produces phase errors. Note however that the integer part of $\tau/2\pi$
is also essential to perform the correct conditional phase shift.
Therefore, in practice it is necessary to specify the duration
of the time evolution $I^{\prime}$ to at least 8 digits (for the case
$|J|/h^z_1\approx10^{-6}$).

\section{Summary}\label{sec:6}

On a physically realizable, non-ideal quantum computer,
operations that manipulate one particular qubit also affect the state of other qubits.
This may cause unwanted deviations from the ideal motion of the total system
and lead to practical problems of programming quantum computers:
An implementation of a quantum computation that works well
on one quantum computer may fail on others.

We have classified the various physical sources that lead to deviations.
The most obvious one originates from the fact that
other spins cannot be kept still during an operation
on one particular spin. If these spins do not return to their
original state when this operation is over, the quantum
computation is unlikely to give correct answers \cite{DERAEDT4}.

Proper optimization of the parameters that control the single-qubit
operations can largely eliminate this source of errors.
However, even if the operation gives almost exact results for all basis states,
the operation is not necessarily perfect.
That is, the operation generally yields a global phase factor which depends on the input states.
Therefore, when such an operation is applied on a linear combination of the
basis states, the relative phases of the basis states change,
resulting in incorrect quantum computation.
This is a second source for deviations from correct quantum computer operation.

We have derived additional conditions on the parameters that control the single-qubit
operations and have obtained the conditions for reliable quantum computation.
Unfortunately, these conditions cannot be satisfied simultaneously.
However they can be satisfied to any precision by increasing the duration
of the single-qubit operations.
Using the controlled-NOT gate and Grover's search algorithm as examples,
we have given concrete demonstrations of how the above mentioned problems arise
and how they can be solved.

At this moment, we do not know how to stabilize
the quantum computation by controlling the evolution of the
state of a closed quantum system.
In a classical computer the presence of dissipation enables reliable computation.
However, dissipation seems detrimental for quantum computer operation.
Therefore, at this moment, the only option is to perform each operation as perfect as possible.
The present paper shows how this may be done.

The condition on the commensurability of the precession frequencies
of the individual qubits leads to an increase of the execution time of single-qubit operations.
Unless the precession frequencies of the qubits
are the same to great precision, the execution time will grow rapidly
with the number of qubits and substantially limit the speed of
quantum computation. Therefore new techniques have to be developed
to compensate for this loss in efficiency.
Quantum error correction schemes that work well on an ideal quantum computer
require many extra qubits and many additional operations to detect and correct errors.
On a physical quantum computer however, the error-correction qubits will suffer from the
same deficiencies as those exposed in this paper.
Possibly, the clever use of dissipation processes may help to perform
automatic error correction \cite{AUTOCORRECT}.
All this puts considerable demands on the technology to fabricate qubits.
It remains a great challenge to demonstrate that a QC of many qubits
can perform a genuine computation in less time than a conventional computer.

\section*{Acknowledgement}
Support from the Dutch ``Stichting Nationale Computer Faciliteiten (NCF)'' and
the Grant-in-Aid from the Japanese Ministry of Education, Science, Sports and Culture
is gratefully acknowledged.

\newpage

\begin{table}[ht]
\caption{Input and output states and the corresponding expectation
values ($a$,$b$) of the qubits for the CNOT operation.}
  \begin{center}
    \begin{tabular}{ccccccccc}
Input state&\vline&$a$&$b$&\vline&Output state&\vline&a&b\\
\hline
$\KET{00}$&\vline &0&0&\vline&$\KET{00}$ &\vline&0&0\\
$\KET{10}$&\vline &1&0&\vline&$\KET{11}$ &\vline&1&1\\
$\KET{01}$&\vline &0&1&\vline&$\KET{01}$ &\vline&0&1\\
$\KET{11}$&\vline &1&1&\vline&$\KET{10}$ &\vline&1&0\\
    \end{tabular}
    \label{tab:CNOT}
  \end{center}
\end{table}

\begin{table}[ht]
    \caption{Model parameters for the elementary operations on the ideal QC.
    Parameters of model (3) that do not
    appear in this table are zero, except for the interaction $J=-0.43\times10^{-6}$.
    The TDSE is solved using a time step $\delta/2\pi=1$.}
  \begin{center}
    \begin{tabular}{cccccccc}
           & $\tau/2\pi$  & $h_{1}^x$ & $h_{2}^x$ & $h_{1}^y$ & $h_{2}^y$ & $h_{1}^z$ & $h_{2}^z$ \\
\hline
$X_1$      &     0.25     &     1       &       0     &      0      &    0        &     0       & 0  \\
$X_2$      &     0.25     &     0       &       1     &      0      &    0        &     0       & 0  \\
$Y_1$      &     0.25    &     0       &       0     &      1      &    0        &     0       & 0  \\
$Y_2$      &     0.25    &     0       &       0     &      0      &    1        &     0       & 0  \\
$X_1^\prime$&     1      &    -0.4477  &       0     &      0      &    0        &     0       & 0  \\
$X_2^\prime$&    1       &     0       & -1.4244     &      0      &    0        &     0       & 0  \\
$Y_1^\prime$&     1      &     0       &       0     &   0.4477    &    0        &     0       & 0  \\
$X_1^{\prime\prime}$&     1      &    -0.6977  &       0     &      0      &    0        &     0       & 0  \\
$X_2^{\prime\prime}$&    1       &     0       & -1.6744     &      0      &    0        &     0       & 0  \\
$I$ & $-1/2J$     &     0       &       0     &      0      &    0        &   $-J/2$ & $-J/2$ \\
$I^\prime$ & $-1/2J$     &     0       &       0  &      0      &    0        &     1       & 0.25   \\
$G$ & $-1/2J$     &     0       &       0  &      0      &    0        &     1       & 0.25   \\
    \end{tabular}
    \label{tab:IDEALQC}
  \end{center}
\end{table}

\begin{table}[ht]
    \caption{Model parameters of single-qubit operations on an NMR QC
    for the case ($k=1$, $N=1$, $M=4$), see
    (\ref{PARAMETERS1}) and (\ref{PARAMETERS2}).
    Parameters of model (3) that do not
    appear in this table are zero, except for the interaction $J=-0.43\times10^{-6}$
    and the constant magnetic fields $h_{1}^z=1$ and $h_{2}^z=0.25$.
    The TDSE is solved using a time step $\delta/2\pi=0.01$.}
  \begin{center}
    \begin{tabular}{ccccccc}
& $\tau/2\pi$ & $\omega$ & $\tilde h_{1}^x$ & $\tilde h_{2}^x$ &  $\tilde h_{1}^y$ & $\tilde h_{2}^y$ \\
\hline
$X_1$ &  8 & 1.00 & 0     & 0 &  -0.0625000 & -0.0156250\\
$X_2$ &128 & 0.25 & 0     & 0 &  -0.0156250 & -0.0039063 \\
$Y_1$ &  8 & 1.00 & 0.0625000 & 0.0156250  & 0 & 0 \\
$Y_2$ &128 & 0.25 & 0.0156250 & 0.0039063  & 0 & 0 \\
$X_1^\prime$ & 8 & 1.00 & 0 & 0 & 0.1119186 & 0.0279796  \\
$X_2^\prime$ &128 & 0.25 & 0 & 0 &0.0890262 & 0.0222565  \\
$Y_1^\prime$ & 8 & 1.00 &       -0.1119186 &-0.0279796 &  0 & 0 \\
$X_1^{\prime\prime}$ & 8 & 1.00 & 0 & 0 & 0.1744186 & 0.0436046  \\
$X_2^{\prime\prime}$ &128 & 0.25 & 0 & 0 &0.1046512 & 0.0261628  \\
    \end{tabular}
    \label{tab:NMRQC}
  \end{center}
\end{table}

\begin{table}[ht]
    \caption{Model parameters of single-qubit operations on an NMR QC using rotating SF's
    for the case ($k=1$, $N=1$, $M=4$), see
    (\ref{PARAMETERS1}) and (\ref{PARAMETERS2}).
    Parameters of model (3) that do not
    appear in this table are zero, except for the interaction $J=-0.43\times10^{-6}$
    and the constant magnetic fields $h_{1}^z=1$ and $h_{2}^z=0.25$.
    The TDSE is solved using a time step $\delta/2\pi=0.01$.}
  \begin{center}
    \begin{tabular}{ccccccccccc}
& $\tau/2\pi$ & $\omega$ & $\tilde h_{1}^x$ & $\tilde h_{2}^x$ & $\varphi_x$ & $\tilde h_{1}^y$ & $\tilde h_{2}^y$ & $\varphi_y$ \\
\hline
$X_1$ &  8 & 1.00        & -0.0312500 & -0.0078125 & $-\pi/2$ & -0.0312500& -0.0078125 & 0 \\
$X_2$ & 128 & 0.25       & -0.0078125 & -0.0039063 & $-\pi/2$ & -0.0078125& -0.0039063 & 0 \\
$Y_1$ &  8 & 1.00        &  0.0312500 &  0.0156250 & 0        &  0.0312500&  0.0156250 & $\pi/2$ \\
$Y_2$ & 128 & 0.25       &  0.0078125 &  0.0039063 & 0        &  0.0078125&  0.0039063 & $\pi/2$ \\
$X_1^\prime$ &  8 & 1.00 &  0.0559593 &  0.0139898 & $-\pi/2$ &  0.0559593&  0.0139898 & 0 \\
$X_2^\prime$ & 128 & 0.25&  0.0445131 &  0.0111283 & $-\pi/2$ &  0.0445131&  0.0111283 & 0 \\
$Y_1^\prime$ &  8 & 1.00 & -0.0559593 & -0.0139898 & 0        & -0.0559593& -0.0139898 & $\pi/2$ \\
$X_1^{\prime\prime}$ &  8 & 1.00 &  0.0872093 &  0.0218023 & $-\pi/2$ &  0.0872093&  0.0218023 & 0 \\
$X_2^{\prime\prime}$ & 128 & 0.25&  0.0523256 &  0.0130914 & $-\pi/2$ &  0.0523256&  0.0130914 & 0 \\
    \end{tabular}
    \label{tab:NMRRFQC}
  \end{center}
\end{table}

\begin{table}[ht]
    \caption{Expectation values of the two qubits as obtained by
    performing a sequence of five CNOT operations on a QC that uses
    rotating SF's to manipulate individual qubits.
    The initial states $\KET{10}$, $\KET{01}$, $\KET{11}$, and
    $\KET{singlet}=(\KET{01}-\KET{10})/\sqrt{2}$ have been prepared by starting from the state
    $\KET{00}$ and performing exact rotations of the spins.
    The CNOT operations on the singlet state are followed by a $\pi/2$ rotation of spin 1 to yield a non-zero
    value of qubit 1.
    The subscripts in $a_{s}$ and $b_{s}$ refer to the time $s=\tau/2\pi=2kMN^2$
    and determine the duration and strength of the
    SF pulses through relations (\ref{PARAMETERS1}) and (\ref{PARAMETERS2}), see
    Table \ref{tab:NMRRFQC} for the example of the case $s=8$.
    The CNOT operation itself was implemented by applying sequence $CNOT_1$ given by (\ref{eq:cnot1}).
    On an ideal QC, CNOT$^4$ is the identity operation.
    The results obtained on an ideal QC are given by $a$ and $b$.}
  \begin{center}
    \begin{tabular}{ccccccccccccccccccc}
Operation&\vline&$a$&$b$&\vline&$a_8$&$b_8$&\vline&$a_{16}$&$b_{16}$&\vline&$a_{32}$&$b_{32}$&\vline&$a_{64}$&$b_{64}$&\vline&$a_{256}$&$b_{256}$ \\
\hline
$(CNOT_1)^5\KET{00}$&\vline           &0.00&0.00&\vline&0.00&0.00&\vline&0.00&0.00&\vline&0.00&0.00&\vline&0.00&0.00&\vline&0.00&0.00\\
$(CNOT_1)^5\KET{10}$&\vline           &1.00&1.00&\vline&1.00&1.00&\vline&1.00&1.00&\vline&1.00&1.00&\vline&1.00&1.00&\vline&1.00&1.00\\
$(CNOT_1)^5\KET{01}$&\vline           &0.00&1.00&\vline&0.00&1.00&\vline&0.00&1.00&\vline&0.00&1.00&\vline&0.00&1.00&\vline&0.00&1.00\\
$(CNOT_1)^5\KET{11}$&\vline           &1.00&0.00&\vline&1.00&0.00&\vline&1.00&0.00&\vline&1.00&0.00&\vline&1.00&0.00&\vline&1.00&0.00\\
$Y_1 (CNOT_1)^5\KET{singlet}$&\vline  &1.00&1.00&\vline&0.90&1.00&\vline&0.03&1.00&\vline&0.58&1.00&\vline&0.88&1.00&\vline&0.99&1.00\\
    \end{tabular}
    \label{tab:NMRRFRESULT1}
  \end{center}
\end{table}

\begin{table}[ht]
    \caption{Same as Table \ref{tab:NMRRFRESULT1} except that instead of $CNOT_1$
    sequence $CNOT_2$ given by (\ref{eq:cnot2}) was used to perform the quantum computation.}
  \begin{center}
    \begin{tabular}{ccccccccccccccccccc}
Operation&\vline&$a$&$b$&\vline&$a_8$&$b_8$&\vline&$a_{16}$&$b_{16}$&\vline&$a_{32}$&$b_{32}$&\vline&$a_{64}$&$b_{64}$&\vline&$a_{256}$&$b_{256}$ \\
\hline
$(CNOT_2)^5\KET{00}$&\vline           &0.00&0.00&\vline&0.24&0.76&\vline&0.50&0.26&\vline&0.20&0.07&\vline&0.06&0.02&\vline&0.00&0.00\\
$(CNOT_2)^5\KET{10}$&\vline           &1.00&1.00&\vline&0.76&0.24&\vline&0.50&0.74&\vline&0.80&0.93&\vline&0.95&0.98&\vline&1.00&1.00\\
$(CNOT_2)^5\KET{01}$&\vline           &0.00&1.00&\vline&0.24&0.24&\vline&0.51&0.74&\vline&0.20&0.93&\vline&0.06&0.98&\vline&0.00&1.00\\
$(CNOT_20^5\KET{11}$&\vline           &1.00&0.00&\vline&0.76&0.76&\vline&0.50&0.26&\vline&0.80&0.07&\vline&0.95&0.02&\vline&1.00&0.00\\
$Y_1 (CNOT_2)^5\KET{singlet}$&\vline  &1.00&1.00&\vline&0.98&0.24&\vline&0.95&0.74&\vline&0.98&0.93&\vline&0.99&0.98&\vline&1.00&1.00\\
    \end{tabular}
    \label{tab:NMRRFRESULT2}
  \end{center}
\end{table}

\begin{table}[ht]
    \caption{Same as Table \ref{tab:NMRRFRESULT1} except that instead of $CNOT_1$
    sequence $CNOT_3$ given by (\ref{eq:cnot3}) was used to perform the quantum computation.}
  \begin{center}
    \begin{tabular}{ccccccccccccccccccc}
Operation&\vline&$a$&$b$&\vline&$a_8$&$b_8$&\vline&$a_{16}$&$b_{16}$&\vline&$a_{32}$&$b_{32}$&\vline&$a_{64}$&$b_{64}$&\vline&$a_{256}$&$b_{256}$ \\
\hline
$(CNOT_3)^5\KET{00}$&\vline           &0.00&0.00&\vline&0.23&0.76&\vline&0.50&0.26&\vline&0.20&0.07&\vline&0.06&0.02&\vline&0.00&0.00\\
$(CNOT_3)^5\KET{10}$&\vline           &1.00&1.00&\vline&0.77&0.24&\vline&0.50&0.74&\vline&0.80&0.93&\vline&0.95&0.98&\vline&1.00&1.00\\
$(CNOT_3)^5\KET{01}$&\vline           &0.00&1.00&\vline&0.23&0.24&\vline&0.51&0.74&\vline&0.20&0.93&\vline&0.06&0.98&\vline&0.00&1.00\\
$(CNOT_3)^5\KET{11}$&\vline           &1.00&0.00&\vline&0.77&0.76&\vline&0.50&0.26&\vline&0.80&0.07&\vline&0.95&0.02&\vline&1.00&0.00\\
$Y_1 (CNOT_3)^5\KET{singlet}$&\vline  &1.00&1.00&\vline&0.79&0.24&\vline&0.55&0.74&\vline&0.82&0.93&\vline&0.95&0.98&\vline&1.00&1.00\\
    \end{tabular}
    \label{tab:NMRRFRESULT3}
  \end{center}
\end{table}

\begin{table}[ht]
    \caption{Same as Table \ref{tab:NMRRFRESULT1} except that instead of rotating SF's, SF's
    along either the $x$ or $y$-axis were used to manipulate individual qubits.
    See Table \ref{tab:NMRQC} for the example of the set of model parameters for $s=8$.}
  \begin{center}
    \begin{tabular}{ccccccccccccccccccc}
Operation&\vline&$a$&$b$&\vline&$a_8$&$b_8$&\vline&$a_{16}$&$b_{16}$&\vline&$a_{32}$&$b_{32}$&\vline&$a_{64}$&$b_{64}$&\vline&$a_{256}$&$b_{256}$ \\
\hline
$(CNOT_1)^5\KET{00}$&\vline           &0.00&0.00&\vline&0.00&0.03&\vline&0.00&0.01&\vline&0.00&0.00&\vline&0.00&0.00&\vline&0.00&0.00\\
$(CNOT_1)^5\KET{10}$&\vline           &1.00&1.00&\vline&1.00&1.00&\vline&1.00&1.00&\vline&1.00&1.00&\vline&1.00&1.00&\vline&1.00&1.00\\
$(CNOT_1)^5\KET{01}$&\vline           &0.00&1.00&\vline&0.00&0.97&\vline&0.00&0.99&\vline&0.00&1.00&\vline&0.00&1.00&\vline&0.00&1.00\\
$(CNOT_1)^5\KET{11}$&\vline           &1.00&0.00&\vline&1.00&0.00&\vline&1.00&0.00&\vline&1.00&0.00&\vline&1.00&0.00&\vline&1.00&0.00\\
$Y_1 (CNOT_1)^5\KET{singlet}$&\vline  &1.00&1.00&\vline&0.02&0.98&\vline&0.45&1.00&\vline&0.17&1.00&\vline&0.70&1.00&\vline&0.98&1.00\\
    \end{tabular}
    \label{tab:NMRRESULT1}
  \end{center}
\end{table}

\begin{table}[ht]
    \caption{Expectation values of the two qubits as obtained by
    running Grover's database search algoritm
    on a QC that uses rotating SF's to manipulate individual qubits.
    The subscripts in $a_{s}$ and $b_{s}$ refer to the time $s=\tau/2\pi=2kMN^2$
    and determine the duration and strength of the
    SF pulses through relations (\ref{PARAMETERS1}) and (\ref{PARAMETERS2}), see
    Table \ref{tab:NMRRFQC} for the example of the case $s=8$.
    The results obtained on an ideal QC are given by $a$ and $b$.}
  \begin{center}
    \begin{tabular}{ccccccccccccccccccc}
Item position&\vline&$a$&$b$&\vline&$a_8$&$b_8$&\vline&$a_{16}$&$b_{16}$&\vline&$a_{32}$&$b_{32}$&\vline&$a_{64}$&$b_{64}$&\vline&$a_{256}$&$b_{256}$ \\
\hline
0&\vline&0.00&0.00&\vline&0.48&0.53&\vline&0.15&0.16&\vline&0.04&0.04&\vline&0.01&0.01&\vline&0.00&0.00\\
1&\vline&1.00&0.00&\vline&0.52&0.50&\vline&0.85&0.15&\vline&0.96&0.04&\vline&0.99&0.01&\vline&1.00&1.00\\
2&\vline&0.00&1.00&\vline&0.55&0.48&\vline&0.15&0.84&\vline&0.04&0.96&\vline&0.01&0.99&\vline&0.00&1.00\\
3&\vline&1.00&1.00&\vline&0.45&0.50&\vline&0.85&0.85&\vline&0.96&0.96&\vline&0.99&0.99&\vline&1.00&0.00\\
    \end{tabular}
    \label{tab:GROVRESULT1}
  \end{center}
\end{table}

\begin{table}[ht]
    \caption{Same as Table \ref{tab:GROVRESULT1} except that instead of rotating SF's,
    SF's along either the $x$ or $y$-axis were used to manipulate individual qubits.
    See Table \ref{tab:NMRQC} for the example of the set of model parameters for $s=8$.}
  \begin{center}
    \begin{tabular}{ccccccccccccccccccc}
Item position&\vline&$a$&$b$&\vline&$a_8$&$b_8$&\vline&$a_{16}$&$b_{16}$&\vline&$a_{32}$&$b_{32}$&\vline&$a_{64}$&$b_{64}$&\vline&$a_{256}$&$b_{256}$ \\
\hline
0&\vline&0.00&0.00&\vline&0.92&0.91&\vline&0.39&0.35&\vline&0.11&0.10&\vline&0.03&0.03&\vline&0.00&0.00\\
1&\vline&1.00&0.00&\vline&0.09&0.91&\vline&0.61&0.36&\vline&0.89&0.10&\vline&0.97&0.03&\vline&1.00&1.00\\
2&\vline&0.00&1.00&\vline&0.95&0.10&\vline&0.36&0.65&\vline&0.10&0.90&\vline&0.03&0.98&\vline&0.00&1.00\\
3&\vline&1.00&1.00&\vline&0.05&0.09&\vline&0.64&0.64&\vline&0.90&0.90&\vline&0.97&0.97&\vline&1.00&0.00\\
    \end{tabular}
    \label{tab:GROVRESULT2}
  \end{center}
\end{table}

\begin{table}[ht]
    \caption{Same as Table \ref{tab:NMRRFRESULT1} except for a change in the
    duration of the operation $I^{\prime}$.
    $(a_{256}^{(1)}$,$b_{256}^{(1)}$): $\tau/2\pi=1162790.4977$;
    $(a_{256}^{(2)}$,$b_{256}^{(2)}$): $\tau/2\pi=1162790.5977$;
    $(a_{256}^{(3)}$,$b_{256}^{(3)}$): $\tau/2\pi=1162790.6977$ (correct value);
    $(a_{256}^{(4)}$,$b_{256}^{(4)}$): $\tau/2\pi=1162790.7977$;
    $(a_{256}^{(5)}$,$b_{256}^{(5)}$): $\tau/2\pi=1162790.8977$.}
  \begin{center}
    \begin{tabular}{ccccccccccccccccccc}
Operation&\vline&$a$&$b$&\vline&$a_{256}^{(1)}$&$b_{256}^{(1)}$&\vline&$a_{256}^{(2)}$&$b_{256}^{(2)}$&\vline&$a_{256}^{(3)}$&$b_{256}^{(3)}$&\vline&$a_{256}^{(4)}$&$b_{256}^{(4)}$&\vline&$a_{256}^{(5)}$&$b_{256}^{(5)}$ \\
\hline
$(CNOT_1)^5\KET{00}$&\vline           &0.00&0.00&\vline&0.00&0.52&\vline&0.00&0.16 &\vline&0.00&0.00 &\vline&0.00&0.13&\vline&0.00&0.48\\
$(CNOT_1)^5\KET{10}$&\vline           &1.00&1.00&\vline&1.00&0.48&\vline&1.00&0.87 &\vline&1.00&1.00 &\vline&1.00&0.84&\vline&1.00&0.48\\
$(CNOT_1)^5\KET{01}$&\vline           &0.00&1.00&\vline&0.00&0.48&\vline&0.00&0.84 &\vline&0.00&0.00 &\vline&0.00&0.87&\vline&0.00&0.52\\
$(CNOT_1)^5\KET{11}$&\vline           &1.00&0.00&\vline&1.00&0.52&\vline&1.00&0.13 &\vline&1.00&1.00 &\vline&1.00&0.16&\vline&1.00&0.52\\
$Y_1 (CNOT_1)^5\KET{singlet}$&\vline  &1.00&1.00&\vline&0.99&0.50&\vline&0.09&0.85 &\vline&0.99&1.00 &\vline&0.01&0.85&\vline&0.99&0.50\\
    \end{tabular}
    \label{tab:NMRRFRESULT4}
  \end{center}
\end{table}

\end{document}